\documentclass[aps,pra,amssymb,amsmath,eqsecnum,nofootinbib]{revtex4}
\usepackage{graphicx}

\newtheorem{definition}{Definition}[section]
\newtheorem{theorem}{Theorem}[section]
\newtheorem{proposition}{Proposition}[section]

\newenvironment{hypothesis}{HP: \begin{center}} {\end{center}}
\newenvironment{thesis}{TH: \begin{center}} {\end{center}}
\newenvironment{proof}{\begin{center}PROOF: \end{center}} {$ \blacksquare $}

\begin{document}
\title{Quantum Democracy Is Possible}
\author{Gavriel Segre}
\begin{abstract}
 It is shown that, since an ultrafilter over an operator-algebraically finite (i.e. isomorphic to the lattice of projections of a finite Von Neumann algebra)  quantum logic  is not necessarily principal, Arrow's Impossibility
Theorem doesn't extend to the quantum case.
\end{abstract}
\maketitle
\newpage
\tableofcontents
\newpage
\section{Acknowledgements}

I would like to thank strongly Vittorio de Alfaro for his
friendship and his moral support, without which I would have
already given up.

Then I would like to thank Piergiorgio Odifreddi for very interest discussions though unfortunately he doesn't agree with this paper.

Finally I would like to thank Giorgio Parisi for very useful suggestions.

Of course nobody among the mentioned people has responsibilities
as to any (eventual) error contained in these pages.
\newpage
\newpage
\section{Introduction}

The differences between Classical Physics and Quantum Physics may
be conceptually interpreted as the fact that the algebraic
probability spaces underlying Classical Physics are commutative
while those underlying Quantum Physics are noncommutative.

Restricting the analysis to the lattice of projections of the
involved Von Neumann algebras  this is equivalent to the fact that
in the classical case such a lattice is Boolean while in the
quantum case it is only orthomodular.

Such a viewpoint constitutes the essence of Quantum Logic, a
research field whose conceptual value in order to understand the
counterintuitive peculiarities of Quantum Mechanics cannot be
overestimated (see \cite{Birkhoff-Von-Neumann-95},
\cite{Beltrametti-Cassinelli-81}, \cite{Cohen-89},
\cite{Svozil-98}, \cite{Ptak-Pulmannova-91},
\cite{Dalla-Chiara-Giuntini-Greechie-04} as to Quantum Logic, see
\cite{Parthasarathy-92}, \cite{Meyer-95} as to Quantum
Probability, and see \cite{Petz-Redei-95},
 \cite{Redei-98},  \cite{Redei-01}, \cite{Ptak-Pulmannova-07},
 \cite{Hamhalter-07} as to the deep link existing between Quantum
 Logic and Quantum Probability).

\bigskip

In a completely different research field, the mathematical
formalization of a democratic voting system led Kenneth Arrow to
prove his celebrated Impossibility Theorem stating that a
perfectly democratic voting system doesn't exists
\cite{Hodge-Klima-05}, \cite{Taylor-05}.

In this paper we show how a lattice theoretic reformulation of
Arrow's Theorem allows to investigate what happens when one
substitutes the underlying classical logic with a quantum one.

We show that, contrary to it classical counterpart, quantum
democracy is possible.
\newpage

\section{Arrow's Impossibility Theorem}
Let us suppose to have an electoral process in which the voters
belonging to a (finite) set V  have to express their preference
among the elements of a (finite) set C of candidates.

Let us recall, with this regard, the following basic:
\begin{definition}
\end{definition}
\emph{partial ordering over C:}

a binary relation $ \succeq $ over C satisfying the following
conditions:
\begin{enumerate}
    \item reflexivity:
\begin{equation}
    c \succeq c \; \; \forall c \in C
\end{equation}
    \item transitivity:
\begin{equation}
 ( c_{1} \succeq c_{2} \: and \: c_{2} \succeq c_{3} \; \Rightarrow
  \;  c_{1} \succeq c_{3} ) \; \; \forall c_{1},c_{2},c_{3} \in C
\end{equation}
    \item identitivity:
\begin{equation}
    ( c_{1} \succeq c_{2} \; and \; c_{2} \succeq c_{1} \;
    \Rightarrow \; c_{1} = c_{2} ) \; \; \forall c_{1} , c_{2} \in C
\end{equation}
\end{enumerate}

\begin{definition}
\end{definition}
\emph{total ordering over C:}

a partial ordering over C such that:
\begin{equation}
    c_{1} \succeq c_{2} \; or \; c_{2} \succeq c_{1} \; \; \forall
    c_{1} , c_{2} \in C
\end{equation}

Let $ \mathcal{O} (C) $ be the set of all the total orderings over
C.

Elections can then be formalized in the following way:
\begin{definition}
\end{definition}
\emph{voting system with voters' set V and candidates' set C:}
\begin{center}
 a map $ S : V \times \mathcal{O}(C) \mapsto \mathcal{O}(C) $
\end{center}

Let $ \mathcal{S}(V,C) $ be the set of all the voting systems with
voters' set V and candidates' set C.

\smallskip

Given $ S \in \mathcal{S}(V,C) $:
\begin{definition}
\end{definition}
\emph{S is democratic:}

it satisfies the following conditions:
\begin{enumerate}
    \item Independence of Irrelevant Alternatives:
\begin{equation}
   ( c_{1} S ( O_{v_{1}} , \cdots , O_{v_{|V|}} )  c_{2}
    \text{ is determined by }  c_{1} O_{v_{1}} c_{2} , \cdots , c_{1} O_{v_{|V|}}
   c_{2} ) \; \; \forall c_{1} , c_{2} \in C , \forall O_{v_{1}} , \cdots ,
   O_{v_{|V|}} \in \mathcal{O}(C)
\end{equation}
    \item Positive Association of Individual Values:
\begin{multline}
    ( c_{1} \succeq_{v_{1}} c_{2} \; and \; c_{1} \succeq_{v_{2}} c_{2} \; and \; c_{1} \succeq_{v_{3}} c_{2} \; and \; c_{1} \succeq_{S|_{ \{ O_{v_{1}} ,  O_{v_{2}} \}
     }} c_{2} \; \Rightarrow \; c_{1} \succeq_{S|_{ \{ O_{v_{1}} ,  O_{v_{2}} , O_{v_{3}}   \}
     }} c_{2} ) \\
     \forall O_{v_{1}},  O_{v_{2}} ,  O_{v_{3}} \in \mathcal{O}(C),   \forall v_{1},v_{2},v_{3} \in
     V , \forall c_{1}, c_{2} \in C
\end{multline}
    \item Citizen Sovereignty:
\begin{equation}
    \nexists c_{1} , c_{2} \in C \; : \; ( c_{1} \succeq_{S ( O_{v_{1}} , \cdots , O_{v_{|V|}} )
    } c_{2} \; \forall O_{v_{1}} , \cdots , O_{v_{|V|}} \in
    \mathcal{O}(C) )
\end{equation}
    \item Nondictatorship:
\begin{equation}
    \nexists v \in V \; : \; ( c_{1} S
    c_{2} \, = \, c_{1} O_{v} c_{2} \; \; \forall c_{1} , c_{2}
    \in C , \forall O_{v} \in  \mathcal{O}(C)  )
\end{equation}
\end{enumerate}

Let $ \mathcal{D}(V,C) $ be the set of all the democratic voting
systems with voters' set V and candidates' set C.

Then:

\begin{theorem} \label{th:Arrow's Impossibility Theorem}
\end{theorem}
\emph{Arrow's Impossibility Theorem:}
\begin{equation}
    \mathcal{D}(V,C)  \; = \; \emptyset \; \; \forall V, C : | V|
    \in \mathbb{N}_{+} \, and \, |C| \in \mathbb{N}_{+}
\end{equation}

\begin{proof}

As it is shown in \cite{Odifreddi-00b}, from a mathematical point of view democratic voting systems are nothing but
non-principal ultrafilters of operator-algebraically finite classical logics,
where the non-principality's constraint implements the Nondictatorship's condition.

Since, as we will see in the theorem \ref{th:About ultrafilters and distributivity} of the next section where the theory of filters over lattices is presented, an ultrafilter over an operator-algebraically finite classical logic is principal the thesis follows.
\end{proof}
\newpage
\section{Quantum democracy}

Let us recall that:

\begin{definition}
\end{definition}
\emph{partially ordered set:}

a couple $ ( S , \succeq ) $ such that:
\begin{enumerate}
    \item S is a set
    \item $ \succeq $ is a partial ordering over S
\end{enumerate}

Given a partially ordered set $  ( S , \succeq ) $:
\begin{definition}
\end{definition}
\emph{meet over  $  ( S , \succeq ) $:}

a map $ \wedge : S \times S \mapsto S $ such that:
\begin{equation}
    x  \; \succeq \; x \wedge y \; \; \forall x, y \in S
\end{equation}
\begin{equation}
    y \;  \succeq \; x \wedge y \; \; \forall x, y \in S
\end{equation}
\begin{equation}
 ( x   \succeq  z \: and \: y \succeq z \;
  \Rightarrow \;  x \wedge y \, \succeq z ) \; \; \forall x, y , z \in S
\end{equation}

\begin{definition}
\end{definition}
\emph{join over  $  ( S , \succeq ) $:}

a map $ \vee : S \times S \mapsto S $ such that:

\begin{equation}
    x \vee y  \; \succeq \; x  \; \; \forall x, y \in S
\end{equation}
\begin{equation}
     x \vee y \;  \succeq \; y \; \; \forall x, y \in S
\end{equation}
\begin{equation}
  (z   \succeq  x \: and \: z \succeq y \;
  \Rightarrow \;  z \, \succeq \, x \vee y ) \; \; \forall x, y , z \in S
\end{equation}

\begin{definition}
\end{definition}
\emph{lattice:}

 $ ( S , \succeq , \wedge , \vee )  $ such that:
 \begin{enumerate}
    \item $  ( S , \succeq ) $ is a partially ordered set
    \item $ \wedge $ is a meet over  $  ( S , \succeq ) $
    \item $ \vee $ is a join over  $  ( S , \succeq ) $
 \end{enumerate}

Given a lattice $ \mathcal{L} := ( S , \succeq , \wedge , \vee ) $
let $ \mathbf{0} $ be its lower bound and let $ \mathbf{1} $ be
its upper bound.

\newpage
\begin{definition}
\end{definition}
\emph{$ \mathcal{L} $ is distributive:}
\begin{equation}
    x \wedge ( y \vee z ) \; = \; ( x \wedge y ) \vee ( x \wedge z) \; \;
    \forall x,y,z \in S
\end{equation}
\begin{equation}
    x \vee ( y \wedge z ) \; = \; ( x \vee y ) \wedge ( x \vee z) \; \;
    \forall x,y,z \in S
\end{equation}

\begin{definition}
\end{definition}
\emph{orthocomplementation over $ \mathcal{L} $:}

a map $ ' : S \mapsto S $ such that:
\begin{equation}
    ( x ' )' \; = \; x \; \; \forall x \in S
\end{equation}
\begin{equation}
    x' \wedge x \; = \;  \mathbf{0}  \; \; \forall x \in S
\end{equation}
\begin{equation}
    x' \vee x \; = \;  \mathbf{1}  \; \; \forall x \in S
\end{equation}
\begin{equation}
    x \preceq y \; \Rightarrow \; y' \preceq x'
\end{equation}

\begin{definition}
\end{definition}
\emph{orthocomplemented lattice:}

a couple $ ( \mathcal{L} , ' ) $ such that:
\begin{enumerate}
    \item $  \mathcal{L} $ is a lattice
    \item $ '  $ is an orthocomplementation over $ \mathcal{L} $
\end{enumerate}

\begin{definition}
\end{definition}
\emph{Boolean lattice:}
\begin{center}
 a distributive orthocomplemented lattice
\end{center}

In the physical literature a Boolean lattice is usually called a
classical logic.
\begin{definition}
\end{definition}
\emph{modular lattice:}

an orthocomplemented lattice $ ( \mathcal{L} , ' ) $ such that:
\begin{equation}
    ( z \succeq x \; \Rightarrow \; ( x \vee y) \wedge z \, = \, x
    \vee ( y \wedge z)) \; \; \forall x,y,z \in L
\end{equation}

\begin{definition}
\end{definition}
\emph{orthomodular lattice:}

an orthocomplemented lattice $ ( \mathcal{L} , ' ) $ such that:
\begin{equation}
   ( y \succeq x \; \Rightarrow \;  x \vee ( x' \wedge y)  \, = \, y ) \; \; \forall x,y \in L
\end{equation}

Let us recall that:

\begin{proposition} \label{prop:hierarchical chain betweem logics}
\end{proposition}
\begin{equation}
    Booleanity \; \Rightarrow \; modularity \; \Rightarrow \; orthomodularity
\end{equation}
\begin{equation}
  orthomodularity \; \nRightarrow \; modularity \; \nRightarrow \;
    Booleanity
\end{equation}

\begin{definition}
\end{definition}
\emph{quantum logic:}
\begin{center}
 a non-Boolean orthomodular lattice
\end{center}

\smallskip

Given two orthocomplemented lattices $ ( \mathcal{L}_{1} ,
\wedge_{1}, \vee_{1} , '_{1} ) $ and $ ( \mathcal{L}_{2} ,
\wedge_{2}, \vee_{2} , '_{2} ) $:

\newpage
\begin{definition}
\end{definition}
\emph{isomorphism of $ ( \mathcal{L}_{1} , \wedge_{1}, \vee_{1} ,
'_{1} ) $ and  $ ( \mathcal{L}_{2} , \wedge_{2}, \vee_{2} , '_{2}
) $:}

a bijective map $ i: \mathcal{L}_{1} \mapsto \mathcal{L}_{2} $
such that:
\begin{equation}
    i ( x \wedge_{1} y ) \; = \; i (x) \wedge_{2} i(y) \; \;
    \forall x,y \in \mathcal{L}_{1}
\end{equation}
\begin{equation}
    i ( x \vee_{1} y ) \; = \; i (x) \vee_{2} i(y) \; \;
    \forall x,y \in \mathcal{L}_{1}
\end{equation}
\begin{equation}
    i( (x)'_{1} ) \; = \; (i(x))'_{2} \; \;
    \forall x \in \mathcal{L}_{1}
\end{equation}

Then:

\begin{theorem} \label{th:structure's theorem about classical logics}
\end{theorem}
\emph{structure's theorem about classical logics:}

\begin{hypothesis}
\end{hypothesis}

\begin{center}
  $ ( \mathcal{L} , \wedge , \vee , ' ) $ Boolean lattice
\end{center}

\begin{thesis}
\end{thesis}
\begin{center}
    $ \exists $ S set such that  $ ( \mathcal{L} , \wedge , \vee , ' ) $
    is isomorphic to $ ( \mathcal{P}(S) , \cap , \cup , - ) $
\end{center}
where $ \mathcal{P}(S) $ is the power set of S and $ - $ denotes
set theoretic complement.

\begin{theorem} \label{th:structure's theorem about orthomodular lattices}
\end{theorem}
\emph{structure's theorem about orthomodular lattices:}

\begin{hypothesis}
\end{hypothesis}

\begin{center}
  $ ( \mathcal{L} , \wedge , \vee , ' ) $ orthomodular lattice
\end{center}

\begin{thesis}
\end{thesis}
\begin{center}
    $ \exists $ A Von Neumann algebra  such that  $ ( \mathcal{L} , \wedge , \vee , ' ) $
    is isomorphic to $ ( \mathcal{P}(A) , \wedge_{A} , \vee_{A} , '_{A} ) $
\end{center}
where $ \mathcal{P}(A) $ is the lattice of projections of A on
which the join $ \vee_{A} $, the meet $ \wedge_{A} $ and the
orthocomplementation $ '_{A} $ are defined in the usual
operator-algebraic way.

Given an orthomodular lattice $ \mathcal{L} $:

\begin{definition} \label{def:finite orthomodular lattice}
\end{definition}
\emph{$\mathcal{L}$ is operator-algebraically finite:}
\begin{center}
  $\mathcal{L}$ is isomorphic to the lattice of projections $ \mathcal{P}(A)
  $ of a finite Von Neumann algebra
\end{center}

\bigskip

Given a subset S of a lattice $ \mathcal{L} $:
\begin{definition}
\end{definition}
\emph{S is upper:}
\begin{equation}
  ( x \in S \, and \, y \succeq x \; \Rightarrow \; y
  \in S ) \; \; \forall x , y \in  \mathcal{L}
\end{equation}

\newpage
\begin{definition}
\end{definition}
\emph{S is lower:}
\begin{equation}
  ( x \in S \, and \, x \succeq y \; \Rightarrow \; y
  \in S )  \; \; \forall x , y \in  \mathcal{L}
\end{equation}

\begin{definition}
\end{definition}
\emph{upper set generated by S:}
\begin{equation}
    S \uparrow \; := \; \{ x \in \mathcal{L} :  (\exists y \in S : x \succeq y )  \}
\end{equation}

\begin{definition}
\end{definition}
\emph{lower set generated by S:}
\begin{equation}
    S \downarrow \; := \; \{ x \in \mathcal{L} :  (\exists y \in S : y \succeq x )  \}
\end{equation}

\begin{definition}
\end{definition}
\emph{S is a filter:}
\begin{equation}
    S  \text{ is upper } \; and \; ( x \wedge y \in S \; \;
    \forall x,y \in S )
\end{equation}

Given a filter F of a lattice $ \mathcal{L} $:
\begin{definition}
\end{definition}
\emph{F is a proper filter:}
\begin{center}
  F is a proper subset of $ \mathcal{L} $
\end{center}

Given a proper filter F of a lattice $ \mathcal{L} $:
\begin{definition}
\end{definition}
\emph{F is a principal filter:}
\begin{equation}
    \exists x \in \mathcal{L}  \; : \; F = \{ x \} \uparrow
\end{equation}

\begin{definition}
\end{definition}
\emph{F is an ultrafilter:}
\begin{equation}
    \nexists F' \; \text{ filter in $ \mathcal{L} $ } \; : \;  F' \supset F
\end{equation}

Then:

\begin{theorem} \label{th:About ultrafilters and distributivity}
\end{theorem}
\emph{About ultrafilters and Booleanity:}

\begin{equation}
   ( F \text{  ultrafilter over an operator-algebraically finite orthomodular lattice $ \mathcal{L} $ } \;
    \Rightarrow \; F \text{ is principal } ) \Leftrightarrow \text{$ \mathcal{L} $ is a classical logic}
\end{equation}

Let us remark that the fact that if $ \mathcal{L} $ is an
operator-algebraically finite quantum logic an ultrafilter is not
necessarily principal may be appreciated considering the following
counterexample: the quantum logic $ \mathcal{P} (R) $ where R is
the hyperfinite $ II_{1} $ factor \cite{Jones-Sunder-97}.

\bigskip

Since, as it is shown in \cite{Odifreddi-00b} and as  we have already  seen in the proof of the theorem \ref{th:Arrow's Impossibility Theorem}, democratic voting systems are nothing but
non-principal ultrafilters of operator-algebraically finite classical logics,
where the non-principality's constraint implements the Nondictatorship's condition,
it appears then natural to introduce the following:

\newpage
\begin{definition}
\end{definition}
\emph{quantum democracy:}
\begin{center}
 a non-principal ultrafilter in a quantum logic
\end{center}

We will denote the set of all quantum democracies as $
\mathcal{Q}\mathcal{D} $.

Then:
\begin{theorem}
\end{theorem}
\emph{Existence theorem of quantum democracy:}
\begin{equation}
  \mathcal{Q}\mathcal{D} \; \neq \; \emptyset
\end{equation}

\begin{proof}

Since a quantum logic is nondistributive, the theorem
\ref{th:About ultrafilters and distributivity} implies that the
fact that a quantum democracy is an ultrafilter over an
operator-algebraically finite orthomodular lattice doesn't imply
that it is principal.

So the Nondictatorship condition is not violated.
\end{proof}

\end{document}